\documentclass[reprint,aps,prl,amsmath,amssymb,nofootinbib,superscriptaddress]{revtex4-1}

\usepackage[T1]{fontenc}
\usepackage[utf8]{inputenc}
\usepackage{graphicx,color,rotating,pifont}
\usepackage{mathtools}
\usepackage{siunitx}
 
\usepackage{bm}
\usepackage{ae}
\usepackage{float}
\usepackage{siunitx}
\usepackage{dcolumn}
\usepackage{txfonts}
\usepackage{tensor}
\usepackage{braket}
\usepackage{booktabs}
\usepackage{xspace}
\usepackage{mathrsfs}
\usepackage{amssymb} 
\usepackage{amsmath}
\usepackage{esvect}
\usepackage{diagbox} 

\usepackage{tabularx}  
\usepackage{soul}
\setstcolor{red}
\setul{}{.2ex} 

\usepackage[normalem]{ulem}

\usepackage{tikz}
\usetikzlibrary{positioning,calc,shapes,decorations.markings,decorations.pathmorphing}
\tikzset{asg/.cd,
  omega-vertex/.style={circle,solid,draw=black,fill=white,minimum size=5pt, inner sep=0pt},
  dbd-vertex/.style={coordinate},
  pline/.style={thick, postaction={decorate}, decoration={markings, mark=at position .5 with {\arrow[xshift=2pt]{stealth}}}},
  hline/.style={thick, postaction={decorate}, decoration={markings, mark=at position .5 with {\arrowreversed[xshift=-2pt]{stealth}}}},
  shift arrow/.style={/pgf/decoration/transform={xshift=#1}},
  shift arrow/.default=-2pt,
  dbd-2b/.style={decorate, decoration=snake},
  omega-2b/.style={densely dashed},
  neutron/.style={draw=blue},
  proton/.style={draw=red},
}

\usepackage[colorlinks,linkcolor=blue,anchorcolor=blue,citecolor=blue]{hyperref} 
\hypersetup{
   colorlinks=true, 
   linkcolor=blue,  
   citecolor=blue,  
   filecolor=blue,  
   urlcolor=blue    
}

\DeclareMathSymbol{\NS}{\mathord}{AMSb}{"4E}

\DeclareSIUnit{\fm}{\femto\meter}

\newcommand{\beq}{\begin{equation}}
\newcommand{\eeq}{\end{equation}}
\newcommand{\beqn}{\begin{eqnarray}}
\newcommand{\eeqn}{\end{eqnarray}}
\newcommand{\bsub}{\begin{subequations}}
\newcommand{\esub}{\end{subequations}}
\newcommand{\bpm}{\begin{pmatrix}}
\newcommand{\epm}{\end{pmatrix}}

\newcommand\identity{1\kern-0.25em\text{l}}

\begin{document} 

\title{Multireference Covariant Density Functional Theory with Stochastic Basis}
 
 \author{X. Zhang}  
    \affiliation{Department of Physics, Kyoto University,
Kyoto 606-8502, Japan}    
  
   \author{K. Hagino}  
   \email{Corresponding author: hagino.kouichi.5m@kyoto-u.ac.jp}
    \affiliation{Department of Physics, Kyoto University,
Kyoto 606-8502, Japan}  
  \affiliation{Institute for Liberal Arts and Sciences, Kyoto University, Kyoto 606-8501, Japan}
\affiliation{ 
RIKEN Nishina Center for Accelerator-based Science, RIKEN, Wako 351-0198, Japan
}

\begin{abstract} 
Multireference density functional theory (MR-DFT) provides a pivotal microscopic framework for the description of 
the ground state properties, low-lying nuclear spectra and transition properties of atomic nuclei. 
Conventionally, practical implementations of MR-DFT rely on empirically chosen generator coordinates, 
which may omit relevant collective degrees of freedom and thus fail to capture sufficient collective correlations. Here we introduce the stochastic-basis multireference density functional theory (MR-SDFT). This is an extended scheme that augments the MR-DFT toolkit by (i) generating a diverse ensemble of mean-field reference configurations via a stochastic external field and (ii) selecting a compact subspace with Projection-Selection method. The chosen reference configurations are then linearly superposed within the MR-DFT framework to yield spectroscopic observables. Applying this framework to \nuclide[20]{Ne}, \nuclide[24]{Mg} and \nuclide[28]{Si} with the 
covariant density functional theory (CDFT), it is demonstrated that the MR-SCDFT leads to lower ground-state energies, smaller point-proton rms radius, and a softer ground-state
band compared to the conventional MR-CDFT. 
\end{abstract}
  
\maketitle

Nuclear collective motions, such as rotations and surface vibrations, play a central role in understanding both the ground-state properties and collective excitations of atomic nuclei — notably clustering phenomenon in light nuclei, fission in heavy nuclei and shape coexistence. Experimentally, these collective modes are manifested through low-lying spectra and characteristic transition patterns, making nuclear spectroscopy the primary window into these phenomena. A microscopic theoretical description of such spectroscopic features is therefore of great importance. Among various microscopic nuclear models, the density functional theory (DFT)~\cite{Bender:2003RMP,Vretenar:2005PR,Meng:2005PPNP,Niksic:2011PPNP,Sheikh:2021qv} has achieved remarkable success in both describing nuclear ground states properties~\cite{Goriely:2009PRL,Erler:2012,Afanasjev:2013,DRHBcMassTable:2024} and low-lying excitations~\cite{Bender:2004Global,Rodriguez:2014Global} of nuclei across almost the whole chart of nuclide.

Although single-reference DFT provides a good description of many nuclear properties, a full description of collective dynamics often requires accounting for elaborate collective correlations and quantum fluctuations beyond a single mean-field configuration~\cite{Yao:2022PPNP}. To include these effects, DFT is commonly combined with quantum-number projections (QNP) and the generator coordinate method (GCM), resulting in the multi-reference DFT (MR-DFT) framework~\cite{Bender:2003RMP,Niksic:2011PPNP,Robledo:2018JPG,Sheikh:2021qv,Yao:2022PPNP}. In MR-DFT,  a many-body wave function is a  superposition of symmetry-restored reference configurations parametrized by generator coordinates, enabling collective correlations 
that are absent in  single reference DFT. This approach  has been proven effective for calculating low-lying spectra~\cite{Bender:2008,Rodriguez:2010PRC,Yao:2010,Niksic:2011PPNP,Bally:2014PRL,Sheikh:2021qv}, cluster phenomena~\cite{Kanada2012PTEP,Zhou:2016}, shape coexistence~\cite{Fu2014PRC,Yang2023PRC}, and nuclear fission~\cite{Goutte2005PRC,Regnier2019PRC,Verriere2021PRC}.

However, conventional MR-DFT calculations typically rely on an empirical selection of only a few generator coordinates (such as deformations and pairing fluctuations), which often fails to fully capture the rich collective correlations in a nuclear many-body system~\cite{Matsumoto2023PRC}. 
As an early attempt, the 
self-consistent collective coordinate (SCC) method was developed to overcome this problem~\cite{Marumori1980PTP,Masayuki1986PTP,Masayuki2000PTP,Nobuo2008PTP,Nakatsukasa2016RMP}. 
The SCC method introduces a time-dependent vibrating frame and determines the collective coordinate by implementing the canonical-variables condition of the collective coordinates and invariance principle of the Schrodinger equation. 
However, since the SCC framework is formulated at the mean-field level, its collective coordinates are not straightforwardly compatible with the GCM framework. Subsequently, several practical methods have been proposed to improve the selection of basis states. These include a stochastic basis-generation approach based on imaginary-time evolution~\cite{Satoshi2006PRC,Fukuoka2013PRC},  and the dynamical GCM~\cite{goeke1980,reinhard1978,Hizawa2021PRC,Hizawa2022PRC}, which introduces the conjugate momentum to the collective coordinate. 
Nevertheless, the former approach may be numerically sensitive to the choice of 
the imaginary-time step size, while the 
latter is computationally demanding, particularly when treating complex many-body correlations such as coupled quadrupole–octupole modes. 

As an alternative method, the optimized-basis GCM has been developed in Refs.~\cite{Matsumoto2023PRC,Matsumoto2025PRC} (see Refs. \cite{myo2023,myo2025} for a similar method for the anti-symmetrized molecular dynamics). This 
method 
performs variational minimization of the total energy with respect to both the single-particle states and the weights of the basis Slater determinants. This approach has been successfully applied to the ground state of $  ^{16}\mathrm{O}  $ and the low-lying states of the sd-shell nuclei $  ^{20}\mathrm{Ne}  $, $  ^{24}\mathrm{Mg}  $, and $  ^{28}\mathrm{Si}  $. 
It was an important finding of these works that one can obtain a better ground state by linearly superposing excited states 
rather than local ground states. 
A drawback of this method, however, is that 
the computational cost remains high  
because the variational optimization must be carried out over a large number of single-particle degrees of freedom, 
especially when full quantum-number projections are required in the MR-DFT framework. 

In this paper, we develop a stochastic-basis multireference DFT (MR-SDFT) approach as an alternative to the optimized-basis GCM. 
In this method, a stochastic external field is added to the single-particle Hamiltonian 
during the self-consistent iteration of the DFT calculations.   
This generates a diverse ensemble of mean-field reference configurations that naturally sample a broad 
region of the collective deformation space. 
The basic idea is that this method simulates the optimization procedure in the optimized-basis GCM, while 
the quantum number projections and the pairing correlations are much more easily implemented. 
Stochastic external fields were originally developed in quantum chemistry~\cite{Mills2017PRA,Ryczko2019PRA} and later introduced into the density functional theory~\cite{Hizawa2023} to provide 
multitudes of 
training data for the deep learning analysis.  A subspace well-represented for low-lying states is then selected 
using the subspace selection method 
~\cite{Zhang2023PRC} based on the energy and orthogonality of the random reference configurations. These selected reference configurations are then linearly superposed by the GCM method with QNP for nuclear spectroscopy.

In this paper, we particularly employ the covariant density functional theory (CDFT). 
The energy functional of CDFT 
consists of the kinetic energy $  \tau(\boldsymbol{r})  $, the nucleon-nucleon interaction energy, and the electromagnetic energy $  \mathcal{E}^{\rm em}(\boldsymbol{r})  $ ~\cite{Burvenich:2002PRC}:
 \beqn
\label{eq:Energy}
E[\tau, \rho, \nabla\rho]
= \int d^3r \Bigl[\tau(\boldsymbol{r})+ \mathcal{E}^{\rm int}(\bm{r}) 
+ \mathcal{E}^{\rm em}(\boldsymbol{r})\Bigr].
\eeqn
Here $  \rho  $ denotes the various densities and the currents constructed as bilinear combinations of the single-particle Dirac wave functions $  \psi_k(\bm{r})  $.
Minimization of the EDF with respect to $\psi^\dagger_k  $ leads to the Dirac equation:
 \beqn
\label{eq:DiracEq}
\left[ \boldsymbol{\alpha} \cdot \mathbf{p} + \beta\left(m + S\right) + V+V_{\rm ext}  \right] \psi_k = \epsilon_k \psi_k ,
\eeqn
which contains the nucleon bare mass $  m  $, the scalar potential $S$, the vector potential $V$, and an external field $  V_{\rm ext}  $. 
$\epsilon_k$ is the single-particle energy, and $\boldsymbol{\alpha}$ and $\beta$ are the Dirac matrices. 
In conventional (shape-constrained) CDFT, $  V_{\rm ext}  $ is chosen to fix the deformation parameters $\beta_\lambda$ to prescribed values, where $\beta_\lambda$ is defined as
 \beqn
 \beta_\lambda = \frac{4\pi}{3AR^\lambda} \braket{\Phi|\hat{Q}_{\lambda 0}|\Phi},
\eeqn
with $  R = 1.2A^{1/3}  $ fm and the multipole moment operators $  \hat{Q}_{\lambda 0} \equiv r^\lambda Y_{\lambda 0}  $. 
Here, $  \ket{\Phi}  $ is the mean-field wave function. 
In the present work, we develop a stochastic CDFT (SCDFT) that generates stochastic mean-field wave functions by replacing the constraint term $  V_{\rm ext}  $ with axial-symmetric, parity-breaking random external fields $  V_{\rm RND}  $~\cite{Hizawa2023}. Each generated mean-field wave function $  \ket{\Phi(\mathbf{\phi})}$ is labeled by $\mathbf{\phi}$, which specifies the realization of the random external field $  V_{\rm RND}  $ used in that calculation. The field $V_{\mathrm{RND}}$ is determined by:
 \beqn
 V_{\mathrm{RND}}(r_\perp,z)=m(r_\perp,z)\,s(r_\perp,z),
 \label{eq:Stochastic_field}
 \eeqn
where the mask function 
 \beqn
 \label{eq:mask}
m(r_\perp,z)=\exp\!\left[-\frac{b}{R^2}\max\left\{0,\sqrt{r_\perp^{2}+z^{2}}-R\right\}^{2}\right],
 \eeqn
confines the field near the nuclear interior and suppresses spurious boundary oscillations. The parameter $b$ is set to be 4 
following Ref. ~\cite{Hizawa2023}. The smoothed field \(s(r_\perp,z)\) in Eq. \eqref{eq:Stochastic_field} is given by 
 \beqn
s(r_\perp,z)=\sum_{r_\perp',z'} s(r_\perp,z;r_\perp',z')\,\nu(r_\perp',z'),
\label{s_function}
 \eeqn
with the Gaussian kernel
   \begin{equation}
s(r_\perp,z;r_\perp',z')=
\exp\!\big[-\big((r_\perp-r_\perp')^{2}+(z-z')^{2}\big)/\mu(r_\perp',z')\big],
\label{Gaussian_kernel}
 \end{equation}
which reduces high-frequency components and produces a smooth, multi-scale random perturbation. At each lattice point \((r_\perp',z')\) we draw an independent uniform random number $\nu(r_\perp',z')$ in the range of $[\nu_{\min}, \nu_{\max}]$ and a random Gaussian width $\mu(r_\perp',z')$ in the range of $[\mu_{\min},\mu_{\max}]$ that avoids biasing the result toward a single length scale.

In MR-DFT, the wave function of low-lying nuclear states is constructed as a superposition of quantum-number projected mean-field wave functions~\cite{Ring:1980},
\begin{equation}
  \ket{\Psi^{JNZ}_\alpha}=\sum^{N_\mathbf{\phi}}_{\mathbf{\phi}} f^{J^\pi}_{\alpha}(\mathbf{\phi}) \ket{JNZ; \mathbf{\phi}},
\end{equation}
 where the index $\alpha$ labels different many-body states that share the same quantum numbers $ J$ and $M$. Here, $J$ is the total angular momentum and $M$ is its projection onto the laboratory $z$-axis. The basis function is constructed using quantum number projections:
 \begin{equation}
 \label{eq:projection}
   \ket{JNZ; \mathbf{\phi}} \equiv  \hat P^J_{M0} \hat P^N\hat P^Z\hat P^\pi\vert \Phi(\mathbf{\phi})\rangle,
 \end{equation}
 where $\hat{P}^\pi,\hat P^Z,\hat{P}^{N},  \hat P^{J}_{M0}$ are the projection operators onto  parity, the proton and the neutron numbers, and the total angular momentum $J$  with the z-component $M$, respectively.  Due to the axial symmetry of the mean-field state $\ket{\Phi(\phi)}  $, only the $  K=0  $ component is nonzero, where $K$ is the projection of the total angular momentum 
 onto the $z$-axis in the intrinsic frame. The weight function $f^{J^\pi}_{\alpha}(\mathbf{\phi})$ is determined with the variational principle which leads to the Hill-Wheeler-Griffin (HWG) equation~\cite{Hill:1953,Ring:1980},
\begin{eqnarray}
\label{eq:HWG}
\sum_{\mathbf{\phi}'}
\Bigg[{\cal H}^{J^\pi}(\mathbf{\phi}, \mathbf{\phi}')
-E_{\alpha}^{J^\pi}{\cal N}^{J^\pi}(\mathbf{\phi}, \mathbf{\phi}') \Bigg] f^{J^\pi}_{\alpha}(\mathbf{\phi}')=0,
\end{eqnarray}
where the Hamiltonian kernel and the norm kernel are defined by
\bsub
\label{eq:GCM_kernel}
\beqn 
      {\cal N}^{J^\pi}(\mathbf{\phi}, \mathbf{\phi}')
     &=& \bra{JNZ; \mathbf{\phi}} JNZ; \mathbf{\phi}'\rangle,\\
     {\cal H}^{J^\pi}(\mathbf{\phi}, \mathbf{\phi}')
    &=& \bra{JNZ; \mathbf{\phi}} \hat H\ket{JNZ; \mathbf{\phi}'},
\eeqn
\esub 
respectively, with the relativistic many-body Hamiltonian, $\hat{H}$. 
The Hamiltonian kernels $ {\cal H}^{J^\pi}(\mathbf{\phi}, \mathbf{\phi}')$ are evaluated with the generalized Wick theorem~\cite{Balian:1969}.

In the following, we take \nuclide[20]{Ne} as an example to demonstrate the feasibility of the present method. Here, we adopt $[\mu_{\min}, \mu_{\max}] = [3.2, 6.4]$ and $[\nu_{\min}, \nu_{\max}] = [-1.6, 1.6]$ for the range of  $\mu$ and $\nu$ in Eqs. (\ref{s_function}) and (\ref{Gaussian_kernel}), respectively. This choice yields the most diverse distribution in multipole-deformation space with a finite number of samples (See Figs.
S1 and S2 in Supplemental Material for a 
comparison of the results with different hyperparameter sets). With this hyperparameter set, we generate an ensemble of
140 mean-field states. We have confirmed that the results do not significantly change even if the number of mean-field states is larger. 
To exclude high energy configurations with excessively large deformations which contribute negligibly to the low-energy nuclear structure, we discard outliers satisfying $\delta E^{0^+}(\phi) \equiv |(E^{0^+}(\phi) - E_g)/E_g| > 0.1$ where $E^{0^+}(\phi)$ is the 
diagonal part of the Hamiltonian kernel for $J^\pi=0^+$ and $E_g$ denotes the global 
ground-state energy obtained from the unconstrained CDFT calculation with the quantum number projections. 
This leaves 114 states out of the 140 mean-field states. 
Here, We use the point-coupling energy functional PC-F1~\cite{Burvenich:2002PRC} due to its better convergence compared to PC-PK1~\cite{Zhao:2010PRC}. 
The large and small components of the single particle wave functions in Eq. \eqref{eq:DiracEq} is expanded on a set of cylinder harmonic oscillator basis with $10$ major shells. The pairing effects are taking into account in the BCS approximation by using a density-independent $\delta$-force with a smooth cut-off~\cite{KRIEGER1990NPA}. 

\begin{figure}[bt]
    \centering 
    \includegraphics[width=\linewidth]{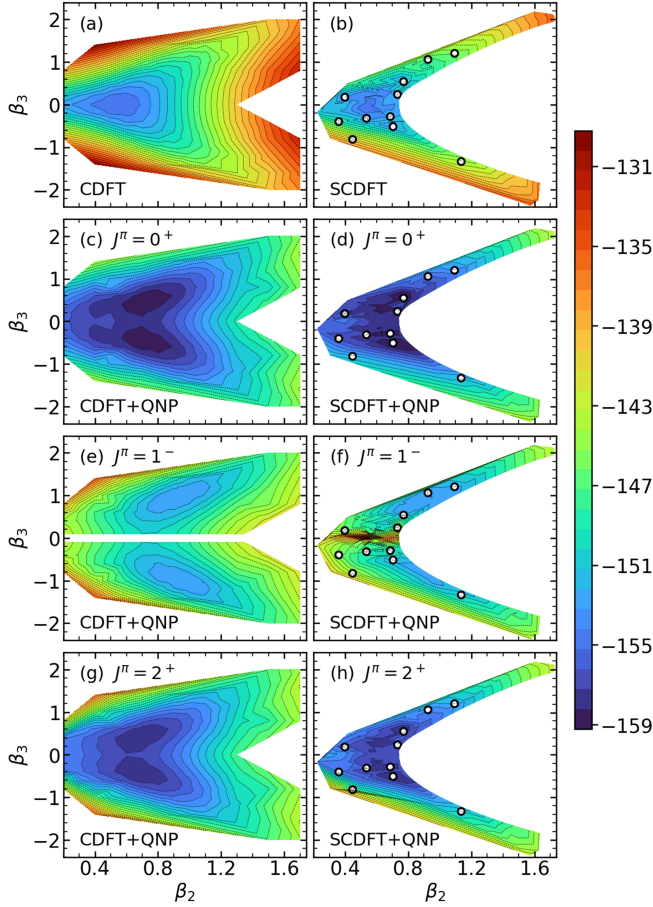}
     \caption{The energy surfaces on the $(\beta_2, \beta_3)$ deformation plane for $^{20}$Ne. Panels (a) and (b) show the energy surfaces for CDFT and SCDFT calculations without projections, respectively. Panels (c), (e), and (g) are the energy surfaces for $J^\pi=0^+$, $J^\pi=1^-$ and $J^\pi=2^+$ from the CDFT+QNP calculations, while Panels (d), (f), and (h) are for the same $J^\pi$ states with the SCDFT+QNP calculations. Neighboring contour lines are separated by 1.0~MeV. The white scatters in the right panels denote the filtered basis states by the Projection-Selection method.}
    \label{fig2}
\end{figure}

Figure~\ref{fig2} presents the energy surfaces on the $(\beta_2, \beta_3)$ plane. 
Figs.~\ref{fig2} (a) and (b) show the energy surfaces  
for the mean-field states with CDFT and the Stochastic CDFT (SCDFT), respectively. 
On the other hand, the other panels show the projected surfaces for $J^\pi = 0^+$, $1^-$, and $2^+$ 
obtained with CDFT+QNP (panels (c), (e), and (g)) and SCDFT+QNP (panels (d), (f), and (h)). 
The full set of energy surfaces up to $ J^\pi = 6^+ $ is presented in Fig. S2 in Supplemental Material.
The locations of the energy minima are similar in both the methods, 
e.g., $(\beta_2,\beta_3)=(0.54,0)$ and (0.56,0.03) in the panels (a) and (b),  
indicating that the dominant quadrupole and octupole deformations are largely preserved. 
However, as is clearly seen in Figs.~\ref{fig2} (a) and (b), the SCDFT+QNP surfaces display visible differences 
and a slightly more complex structure compared with the smoother CDFT+QNP surfaces. 
Specifically, there is a high energy ridge around $\beta_2<0.8, \, \beta_3\approx 0$ for the $J^\pi =1^-$ energy surface in Fig.~\ref{fig2} (f). This disparity arises because parity-odd states cannot gain octupole correlation energy when $\beta_3$ is vanishingly small; their odd-parity content must instead be supplied by higher-order odd multipoles (primarily $\beta_5$ and above) with less coupling to $\beta_2$ than $\beta_3$. These differences arise from the inclusion of higher-order multipole deformations that are naturally generated by the stochastic external field but are absent in conventional constrained CDFT calculations.

It is computationally demanding to evaluate the Hamiltonian kernels in Eq. \eqref{eq:GCM_kernel} if 
all the generated mean-field configurations are included in the MR-CDFT calculations. 
To resolve this problem, we employ a configuration selection algorithm to identify a well-performing subspace. Specifically, we employ the Projection-Selection (PS) 
method~\cite{Zhang2023PRC}\footnote{This method was originally referred to as 
the Orthogonality Condition (OC) Method in Ref. \cite{Zhang2023PRC}. We avoid using this name as it might induce a confusion with 
the Orthogonality Condition Method (OCM) \cite{saito1969}
developed in the cluster physics.}, which selects an optimal subspace using only the norm kernels and the diagonal elements of the Hamiltonian kernels for the desired angular momentum $  J  $. The detailed selection procedure is explained in the Supplemental Material. Applying PS to the $0^+$ state yields a selected subspace of 11 configurations, whose $(\beta_2, \beta_3)$ deformations are depicted with the white scatters in Fig.~\ref{fig2}. 
The SCDFT+QNP method combined with the PS method generates the representative reference configurations in the stochastic configuration space, and the whole procedure is called the stochastic-basis multireference CDFT (MR-SCDFT).

\begin{figure}[bt]
    \centering 
\includegraphics[width=1\linewidth]{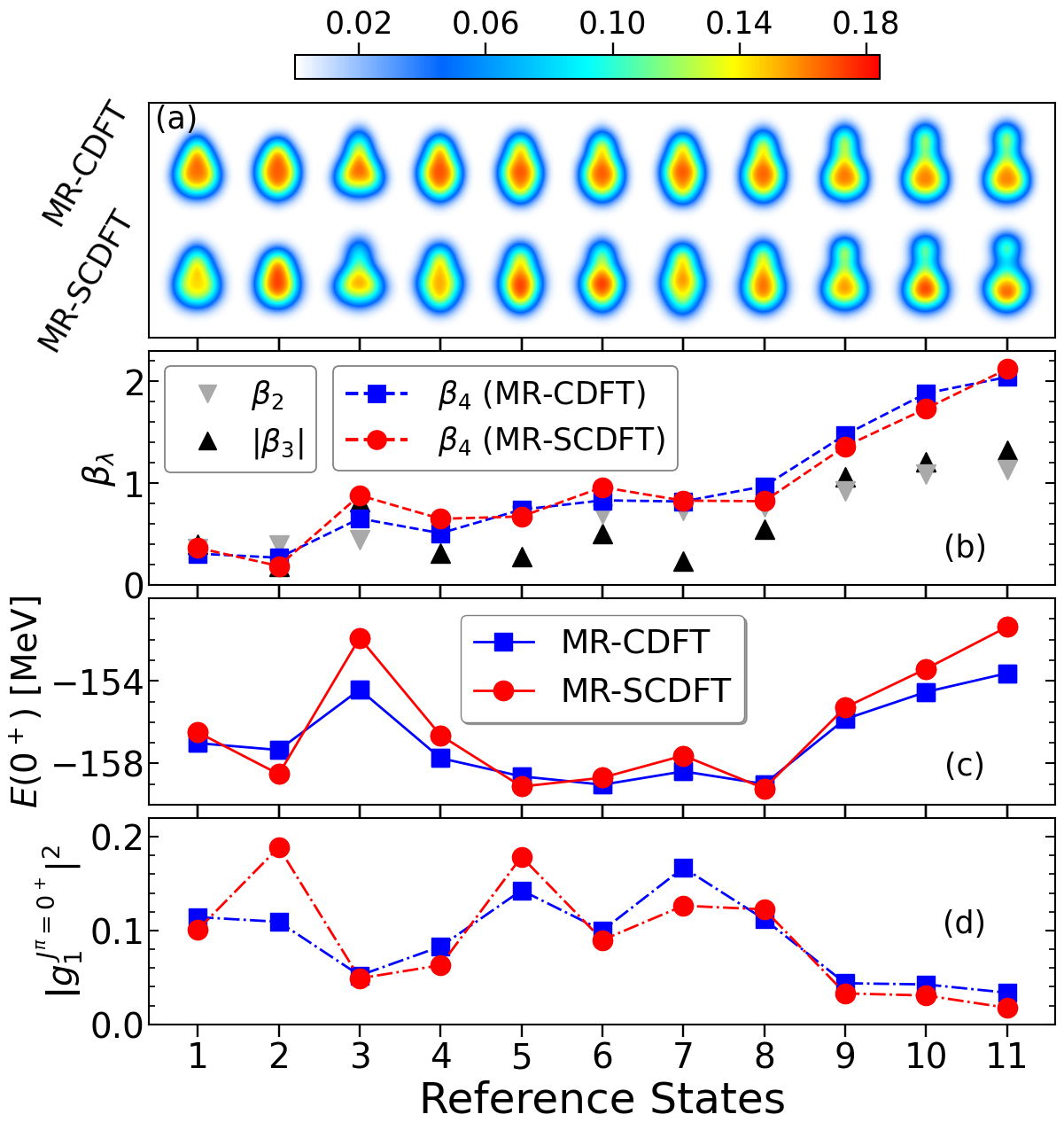}
    \caption{Properties of the eleven reference configurations used in the MR-SCDFT calculations. 
    The corresponding configurations for the MR-CDFT calculations, obtained by constraining to the same  
     $  (\beta_2, |\beta_3|)  $, are also shown. 
    (a) The single-particle density distributions in units of fm$^{-3}$ for MR-CDFT (the upper row) and MR-SCDFT (the lower row); (b) The deformation parameters of $  \beta_2  $ (the gray downward triangles) and  $  |\beta_3|  $ (the black upward triangles), together with  $  \beta_4  $ (the blue squares with dashed line for MR-CDFT and the red circles with dashed line for MR-SCDFT); (c) The projected energies for $J^\pi=0^+$; (d) The distribution of collective wave functions $  |g_1^{J^\pi=0^+}|^2  $. }
\label{fig3}
\end{figure}

Figure~\ref{fig3} (a) shows the resultant one-body density of each configuration for $^{20}$Ne, while Figs.~\ref{fig3} (b-d) show the properties of the reference configurations for MR-SCDFT (the red circles) and MR-CDFT (the blue squares) with the constraint on the same quadrupole and octupole
deformations. Even though the overall shapes of the single-particle densities obtained from MR-CDFT and MR-SCDFT appear similar, 
small differences in the higher-order deformation parameters such as $\beta_4$ shown in Fig.~\ref{fig3} (b) lead to 
significant differences in the projected ground-state energies, with deviations reaching up to 3 MeV between the two 
methods as shown in Fig.~\ref{fig3} (c). 
Notice that, because we employ the 
variation-before-projection (VBP) scheme, 
the projected SCDFT+QNP energy (the red symbols) sometime becomes lower than the corresponding CDFT+QNP energy (the blue symbols), 
even though the CDFT energies should always be lower than the SCDFT energies for the same $(\beta_2,\beta_3)$. 
Furthermore, Fig.~\ref{fig3} (d) displays the distribution of the ground-state wave function $g_{\alpha}^{J^\pi}(\phi) \equiv \sum_{\phi'} \left[ \mathcal{N}^{J^\pi}\right]^{1/2} ({\phi,\phi'}) f_{\alpha}^{J^\pi }(\phi')$. The distribution of MR-SCDFT basis exhibits a clear shift in its peak position: a new dominant peak emerges at the second reference configuration with ($\beta_2, |\beta_3|)\approx$  (0.40, 0.18), in sharp contrast to the MR-CDFT result, where the main peak 
is located 
at the 7-th reference configuration with ($\beta_2, |\beta_3|)\approx$ (0.73, 0.24). This highlights a fundamental difference in how the two approaches describe the collective wave functions.
For the negative-parity states, the dominant contributions arise from the 9-th configuration with ($\beta_2, |\beta_3|)\approx$ (0.92, 1.07), which is consistent with Ref.~\cite{Zhou:2016}. 
(See Figs. S4 and S5 in Supplemental Material for the projected energies and the wave-function distributions 
for up to $J^\pi=6^+$, including negative parity).

\begin{table}
\tabcolsep=4.6pt
\renewcommand{\arraystretch}{1.5}
\caption{The ground state energy, the point-proton root-mean-square (rms) radius, the ratios of the 
excitation energies: $ R_{4/2} = E_x(4^+_1)/E_x(2^+_1) $
and $R_{6/4} = E_x(6^+_1)/E_x(4^+_1)$, and the $E2$ transition strengths from $2_1^+$ to $0_1^+$ of \nuclide[20]{Ne} obtained with MR-CDFT and MR-SCDFT calculations. 
These are compared with the experimental data taken from Refs. ~\cite{NNDC,ANGELI201369}. 
}

\begin{tabular}{llllll}
\hline
\hline
 & $E_g$[MeV]&$r_p$[fm] & $R_{4/2}$ & $R_{6/4}$& $B(E2)$[$e^2\mathrm{fm}^4$]\\
\hline
Exp. &$-$160.645 &2.889(2) & 2.600 & 2.066&65.47(3.22)\\
MR-CDFT &$-$159.452  &3.000 & 3.217 &2.112&66.68\\
MR-SCDFT  &$-$160.511 &2.931 & 2.825 &2.044&61.46 \\ 
\hline\hline
\end{tabular}
\label{tab1}
\end{table}

\begin{figure*}[t]
    \centering 
    \includegraphics[width=2\columnwidth]{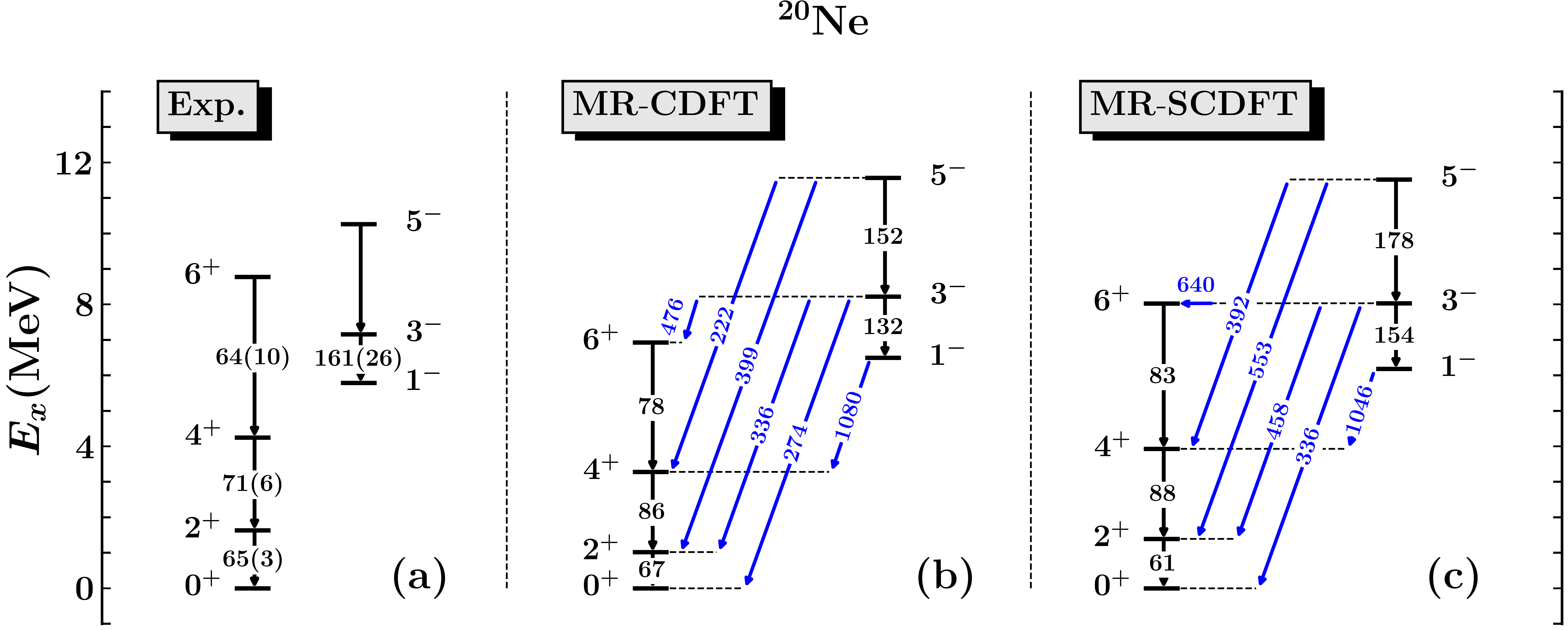}
    \caption{The spectroscopy of \nuclide[20]{Ne} obtained with MR-CDFT and MR-SCDFT calculations in comparison with the 
    available data~\cite{NNDC}.  The $E2$ ($E3$) transitions are denoted by the black (blue) lines, 
    with the corresponding transition strengths given in units of $e^2\mathrm{fm}^4$ and $e^2\mathrm{fm}^6$, respectively.}
    \label{fig4}
\end{figure*}

The ground state energy, the point-proton rms radius, and the ratios of the excitation energies $ R_{4/2} = E_x(4^+_1)/E_x(2^+_1) $
and $R_{6/4} = E_x(6^+_1)/E_x(4^+_1)$ of MR-SCDFT calculation for \nuclide[20]{Ne} are summarized 
in Table~\ref{tab1}. For benchmarking, 
we also perform MR-CDFT calculations on uniformly sampled points on the $(\beta_2, \beta_3)$ plane, mixing the same number 
of configurations to provide a direct comparison. 
Notice that this is different from the MR-CDFT calculations shown in Fig.~\ref{fig3}, in which 
$\beta_2$ and $\beta_3$ are set to be the same as the configurations for MR-SCDFT. 
As one can see in the Table, 
the MR-SCDFT ground-state energy is lower than that from MR-CDFT by 1.1 MeV. 
This indicates  
that the stochastically selected configurations in MR-SCDFT 
form a more complete and well-performing basis, capturing additional correlations compared with the empirical choice of the configurations~\cite{Matsumoto2023PRC,Matsumoto2025PRC}. 
One can also notice that 
the experimental data for the proton rms radius $  r_p  $ and for the $R_{4/2}$ and $R_{6/4}$ ratios are all reproduced better 
with MR-SCDFT as compared to MR-CDFT, even though $B(E2;2_1^+\!\to\! 0^+_1) $ is somewhat underestimated. 
Moreover, the $R_{4/2}$ and the $R_{6/4}$ ratios predicted by MR-SCDFT are systematically smaller than those 
obtained with MR-CDFT, indicating a deviation from a rigid rotor-like structure, 
leading to a reduced rotational collectivity in the low-lying spectra.  
Furthermore, the corresponding ratio $  R_{5/3}=[E_x(5^-_1)-E_x(1^-_1)]/[E_x(3^-_1)-E_x(1^-_1)] $
in the $  K^\pi=0^-  $ band exhibit similar trends, 
but with smaller differences between MR-CDFT  and  MR-SCDFT compared with the differences in the first $  K^\pi=0^+  $ 
band (see Table. S2 in Supplemental Material).

Figure~\ref{fig4} compares the low-lying spectra of the ground-state $  K^\pi=0^+  $ and $ 0^-  $  bands in $^{20}$Ne 
obtained with MR-CDFT and MR-SCDFT, in comparison to the experimental data. 
The excited $K^\pi=0^+$ bands in $^{20}$Ne, which are dominated by strong $  \alpha  $-clustering~\cite{Nauruzbayev2017PRC,Chiba2016PRC}, lie outside the scope of the present framework and are not considered here. Overall, MR-SCDFT provides an improved description of the low-lying spectrum compared to MR-CDFT, particularly in the band-head energy and the level spacing of the ground-state band. 
The $B(E3)$ and $B(E2)$ values of the $K^\pi=0^-$ band in Figs.~\ref{fig4}(b) and (c) exhibit larger relative differences than the 
$B(E2)$ values for the $K^\pi=0^+$ band, which may be due to the contribution from the effect of the higher-order 
odd multipole deformations.

We apply the same methodology to $^{24}$Mg and $^{28}$Si, with selected subspaces of 12 and 15 configurations, respectively. 
The results are shown in Supplemental Material: 
they exhibit the similar overall trend as those for \(^{20}\)Ne. That is, MR-SCDFT predicts lower ground-state energies, smaller point-proton rms radii, and smaller \(R_{4/2}\) and \(R_{6/4}\) ratios than MR-CDFT. 
Notice that the result for the \(R_{4/2}\) ratios is somewhat 
different from that with the optimized GCM reported in Ref.~\cite{Matsumoto2025PRC}, where the \(R_{4/2}\) ratios 
of these three nuclei remain nearly unchanged  under the optimization scheme. 
The corresponding low-lying spectra of \nuclide[24]{Mg} and \nuclide[28]{Si} are shown in Fig. S6 in Supplemental Material. 
For all the three nuclei, MR-SCDFT consistently outperforms MR-CDFT in reproducing the experimental features, particularly by enlarging the underestimated spacing of the first $  K^\pi=0^+  $ band and reducing the overestimation of the $  K^\pi=0^-  $ 
band-head energies of \nuclide[20]{Ne} and \nuclide[24]{Mg}. 
Notably, in \nuclide[28]{Si}, MR-SCDFT significantly improves the description of the third prolate shape-coexistence $ K^\pi=0^+$ band~\cite{Frycz2024PRC} by capturing the rotational-like level spacing and substantially 
enhancing the $B(E2; 4_3^+\!\to\! 2_3^+)$ value.

In summary, we have developed the MR-SCDFT method 
for nuclear low-lying states, by introducing stochastic external fields 
to single-particle Dirac Hamiltonians. The resultant many-body wave functions were then  
filtered to the effective subspace with the Projection-Selection (PS) method before they were linearly superposed. 
We have demonstrated that the stochastic external fields can generate a diverse ensemble of mean-field configurations 
that naturally sample a much broader region of the multidimensional deformation space than with the conventional approach.
We have applied MR-SCDFT to the sd-shell nuclei $^{20}$Ne, $^{24}$Mg, and $^{28}$Si, and 
consistently obtained lower ground state energies by
more than 1 MeV, smaller point-proton rms radii and a less rigid ground-state band, as well as significantly improved low-lying excitation bands.

The present method is readily extendable to heavier nuclei and also with inclusion of non-axial stochastic external fields. 
Moreover, the method can be straightforwardly extended to other functionals than the covariant density functional employed in this paper. Future applications will focus on systematic studies across the nuclear chart and on further refinements of the stochastic sampling protocol to achieve even higher spectroscopic precision.

\section*{Acknowledgments} 
 We thank J. M. Yao, M. Matsumoto, Y. Tanimura and K. Uzawa for fruitful discussions. This work was supported by JST SPRING, Grant Number JPMJSP2110 and by JSPS KAKENHI Grant Number JP23K03414. 
The numerical calculations were performed with the computer facility at the Yukawa Institute for Theoretical Physics, Kyoto University, and the RCNP Computational Facility at Research Center for Nuclear Physics, Osaka University.

\bibliographystyle{apsrev4-1} 
 \bibliography{ref}

\end{document}


\newcommand{\BEhead}[2]{%
\begin{tabular}[c]{@{}c@{}}
$B(E2;$\\
$#1\!\to\!#2)$
\end{tabular}}

\begin{frontmatter} 
\title{Supplemental Material} 
\end{frontmatter}


\setcounter{section}{0}
\setcounter{equation}{0}
\setcounter{figure}{0}
\setcounter{table}{0}

\renewcommand{\theequation}{S\arabic{equation}}
\renewcommand{\thefigure}{S\arabic{figure}}
\renewcommand{\thetable}{S\arabic{table}}


\section{Selection criteria for hyperparameters}

\begin{figure}[bt]
    \centering 
    \includegraphics[width=\columnwidth]{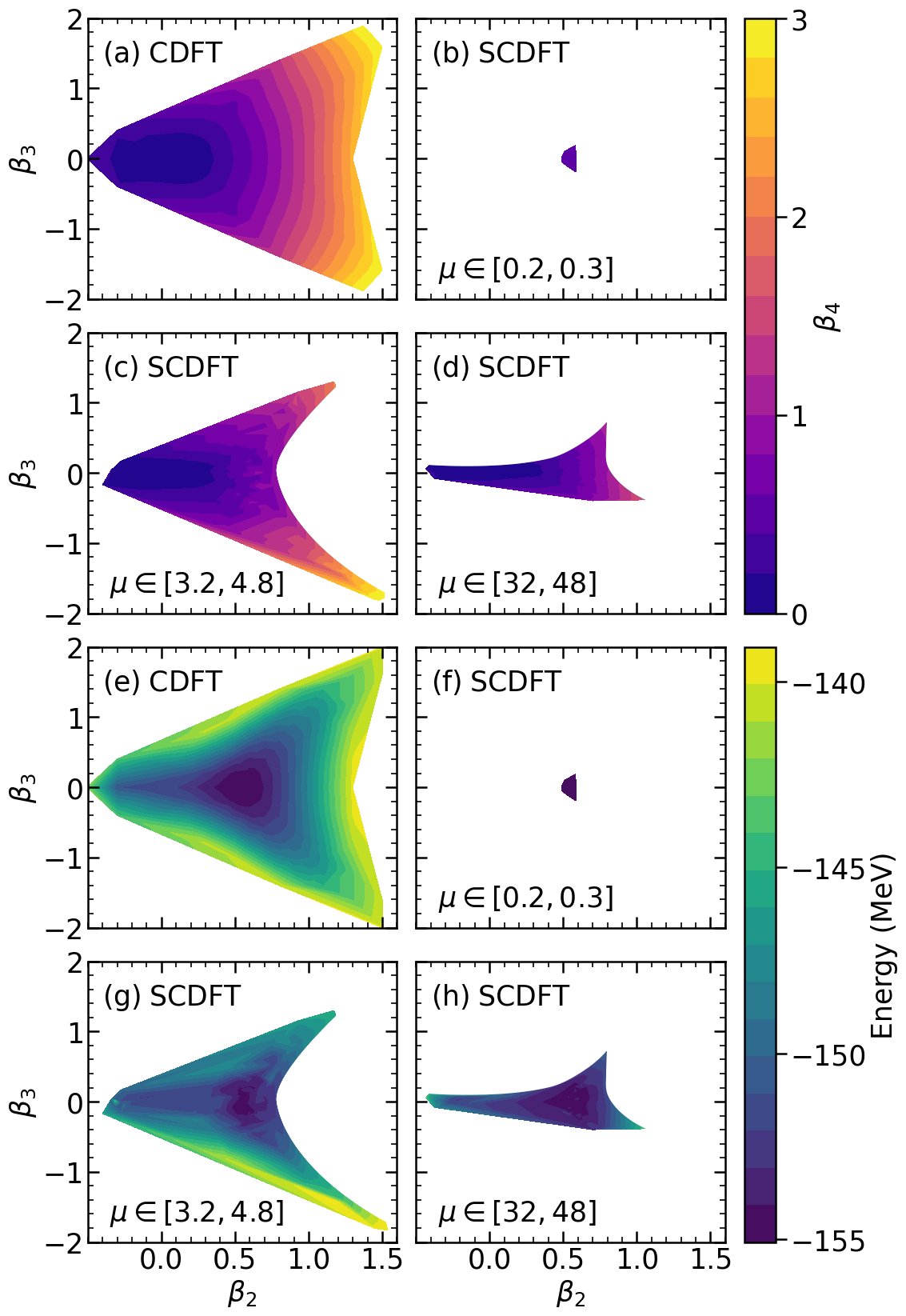}
    \caption{ Mean-field surfaces for $\beta_4$ deformation (the panels (a)-(d)) and for 
    the total energy (the panels (e)-(h)) on the $(\beta_2,\beta_3)$ plane for \nuclide[20]{Ne}. 
    The 
    panels (a) and (e) show the results for CDFT with the shape constraint to $(\beta_2,\beta_3)$, while the others correspond to SCDFT with different choices of the range of $\mu$, that is, $\mu \in [0.2,0.3]$ (the panels (b) and (f)), $\mu \in[3.2,4.8]$  (the panels (c) and (g)), and $\mu \in[32,48]$ 
    (the panels (d) and (h)) with $[\nu_{\rm min}, \nu_{\rm max}]$ fixed to $[-1.6,1.6]$.}
    \label{fig0}
\end{figure}

\begin{figure}[bt]
    \centering 
    \includegraphics[width=\columnwidth]{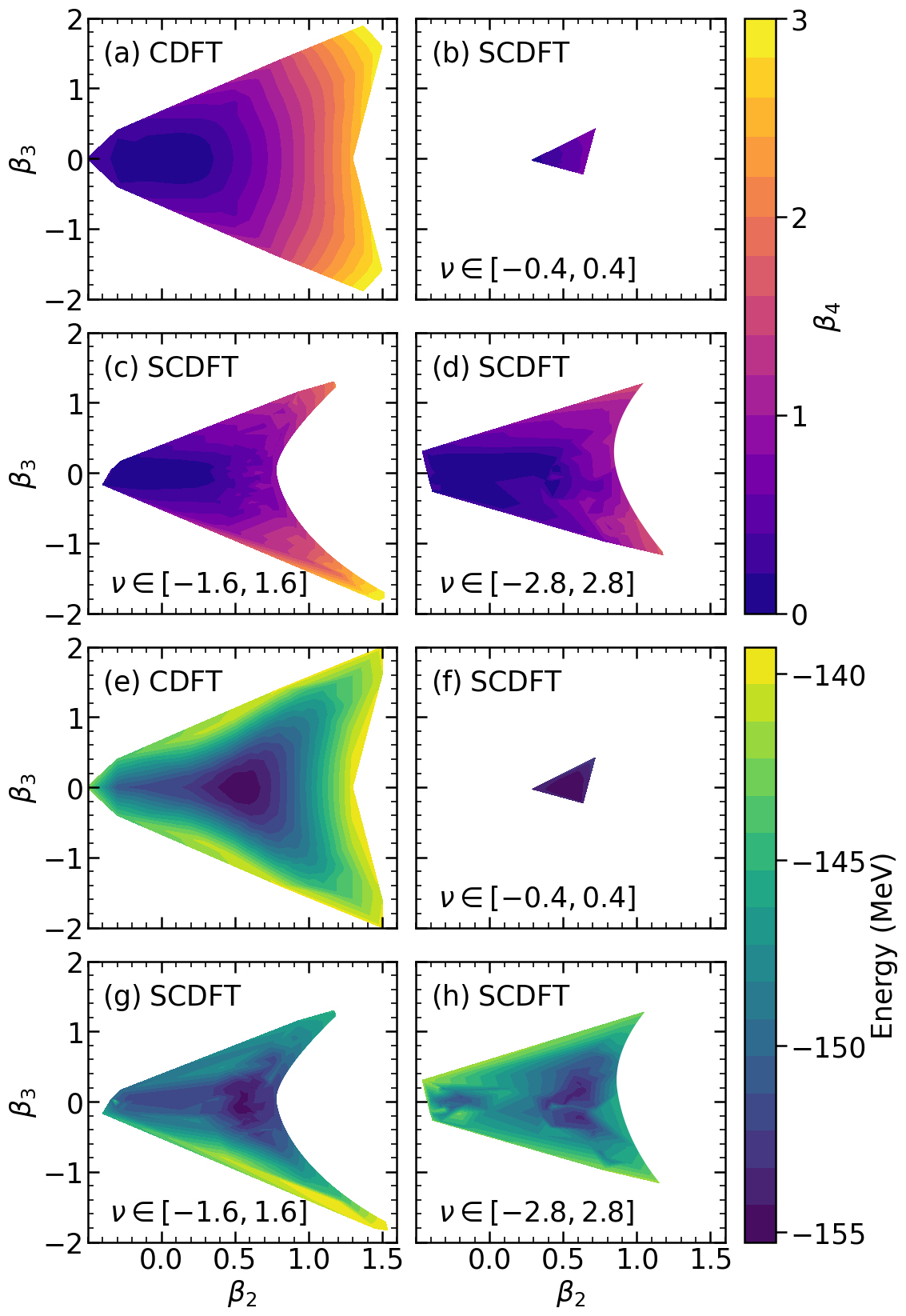}
    \caption{ The same as Fig. \ref{fig0}, but for different choices of  $[\nu_{\rm min}, \nu_{\rm max}]$ as indicated in the figure 
    for a fixed range of $[\mu_{\rm min}, \mu_{\rm max}]=[3.2,4.8]$.}
    \label{fig-1}
\end{figure}

Here we discuss how we select the hyperparameters for the stochastic 
external fields. For this purpose, 
Fig.~\ref{fig0} shows the $  \beta_4  $ and the energy surfaces on the $  (\beta_2, \beta_3)  $ plane for $  ^{20}\mathrm{Ne}  $ obtained with CDFT (the panels (a) and (e), respectively), in comparison with those 
of SCDFT with different sets of the hyperparameters (the panels (b), (c), (d) for $\beta_4$  and (f), (g), (h) for energies). 
In the CDFT calculations, the $(\beta_2, \beta_3) $ plane is uniformly discretized 
to carry out constrained CDFT calculations. 
For SCDFT, we fix  $[\nu_{\min}, \nu_{\max}]$ at $[-1.6, 1.6]$ and 
generate 140 mean-field states for each choice of $[\mu_{\min}, \mu_{\max}]$. 
It is well known that the mean-field surfaces are symmetric about 
$\beta_3=0$ as a consequence of the parity conservation of the energy functional, as shown in Figs.~\ref{fig0} (a) and (e). In contrast, the added stochastic field disrupts this symmetry within finite sample-states as can be seen in 
Figs.~\ref{fig0} (b), (c), and (d) as well as (f), (g), and (h). 
Moreover, we notice that 
the choice of $\mu \in [3.2, 4.8]$ leads to 
the broadest span in both the deformation parameters and the energy among the three choices shown 
in the figure. 
We have confirmed that 
this superiority of the $  [3.2, 4.8]  $ interval for $  \mu  $ remains robust across 
different choices of $  [\nu_{\min}, \nu_{\max}] $. 
After we fix this range for $\mu$, we investigate the range for $\nu$. As shown in Fig. \ref{fig-1}, the choice of $\nu \in [-1.6, 1.6]$ leads to the broadest span among the three choices shown in the figure.  We thus adopt these ranges of the hyperparameters in the 
calculations shown in the main text. 

\section{Additional results for $^{20}$Ne}

\begin{figure}[t]
\centering
\includegraphics[width=\columnwidth]{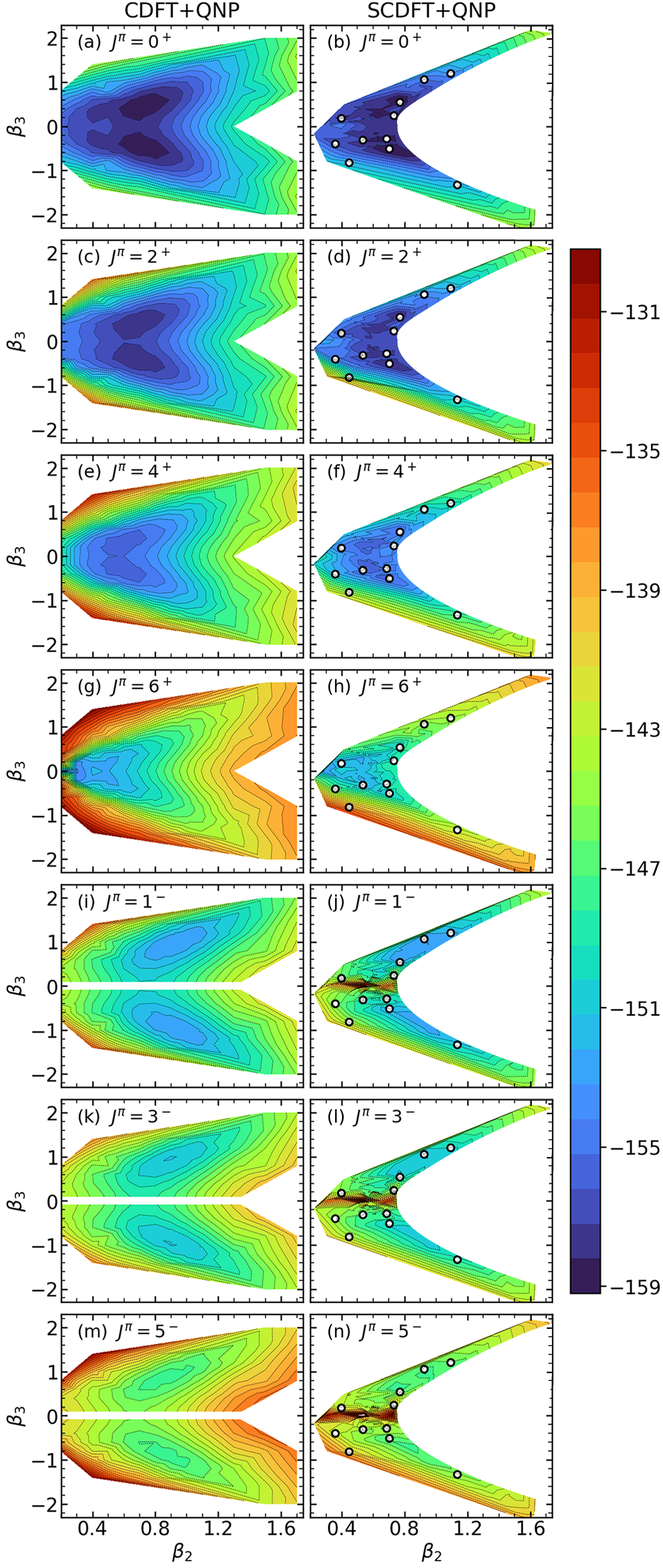}
\caption{Same as Fig. 3 in the main text, but showing the energy surfaces for up to $6^+$ 
states of $^{20}$Ne. }
\label{fig1}
\end{figure}

\begin{figure}[t]
\centering
\includegraphics[width=\columnwidth]{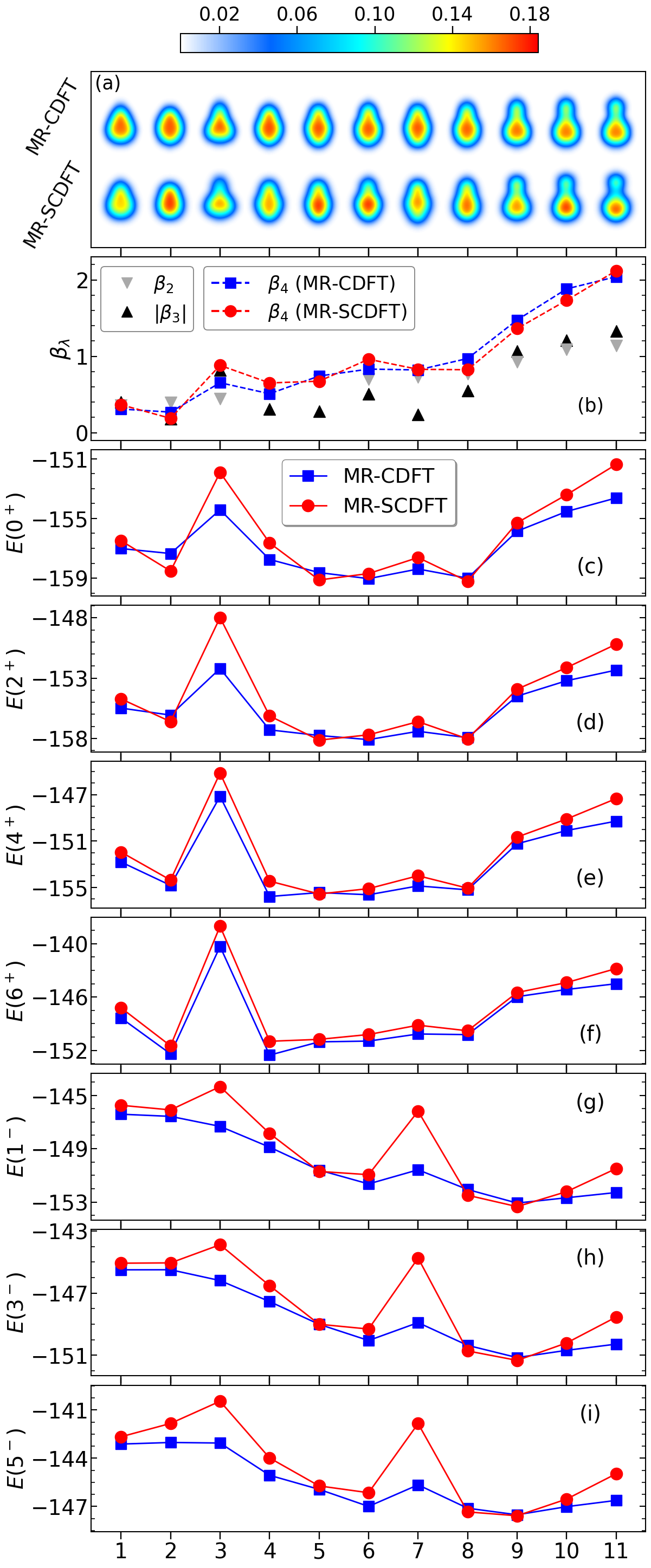}
\caption{Panels (a) and (b) are the same as those in Fig. 2 in the main text, 
while the panels (c)-(i) show the projected energies 
for states up to $J^\pi=6^+$. 
}
\label{fig2}
\end{figure}

\begin{figure}[t]
\centering
\includegraphics[width=\columnwidth]{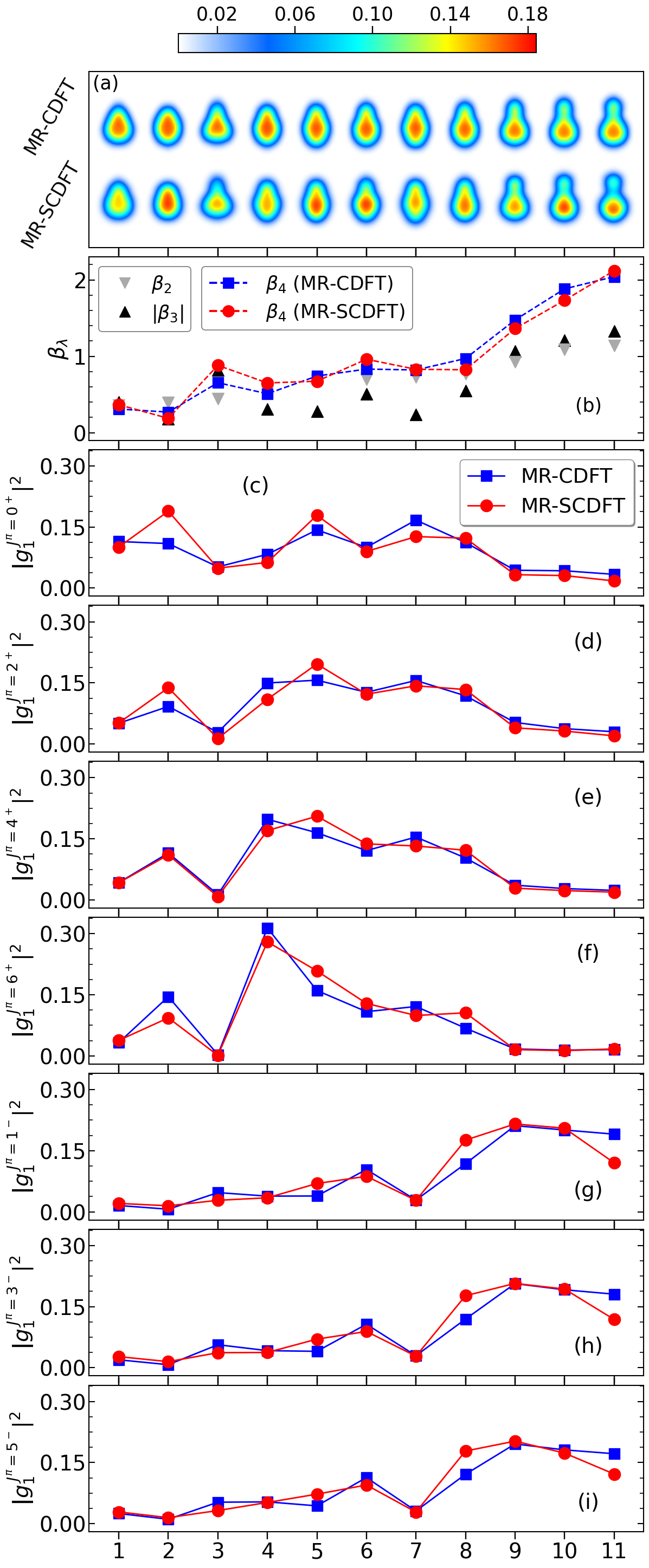}
\caption{Same as Fig. \ref{fig2}, but for the collective wave functions. }
\label{fig3}
\end{figure}

In this section, we show the projected energy surfaces, the projected energies for 
the selected basis configurations, and the collective wave functions for 0$^+$, 1$^-$, 2$^+$, 
$3^-$, 4$^+$, 5$^-$, and 6$^+$ states of $^{20}$Ne. 
These are shown in Figs. \ref{fig1}, \ref{fig2}, and \ref{fig3}, respectively.

\section{Additional results for sd shell nuclei \nuclide[24]{Mg} and \nuclide[28]{Si}}

\begin{table}[t]
\renewcommand{\arraystretch}{1.5}
\caption{Same as Table 1 in the main text, but also with $  ^{24}\mathrm{Mg}  $ 
and $  ^{28}\mathrm{Si}  $. The table includes also the ratio $  R_{5/3} = [E_x(5^-_1) - E_x(1^-_1)] / [E_x(3^-_1) - E_x(1^-_1)]  $.}
\resizebox{\columnwidth}{!}{
\begin{tabular}{llllll}
\hline
\hline
 & $E_g$(MeV)&$r_p$(fm) & $R_{4/2}$ & $R_{6/4}$&$R_{5/3}$\\
\hline
$^{20}$Ne \\ 
Exp. &-160.645 &2.889(2) & 2.600 & 2.066&3.269\\
MR-CDFT &-159.452  &3.000 & 3.217 &2.112&2.934\\
MR-SCDFT  &-160.511 &2.931 & 2.825 &2.044&2.889\\ 
\hline
$^{24}$Mg \\
Exp. &-198.257 &2.941(2) & 3.012 &1.968&3.080 \\
MR-CDFT &-197.273&3.032 & 2.988 & 2.068&2.910\\
MR-SCDFT  &-198.684&3.010 & 2.762 &1.985&2.821\\ 
\hline
$^{28}$Si \\
Exp. &-236.537 &3.010(3)&2.596 &1.850 &-\\
MR-CDFT &-235.223 &3.113& 3.184 &2.122&-\\
MR-SCDFT  &-236.256 &3.062& 2.606 &1.951 &-\\ 
\hline\hline
\end{tabular}}
\label{table1}
\end{table}

\begin{table}[t]
\renewcommand{\arraystretch}{1.5}
\caption{The values of the transition strengths, 
$ B(E2;4^+\!\to\! 2^+)$, $B(E2;2^+ \!\to\! 0^+) $, $B(E2;5^-\!\to\! 3^-)$ and $B(E2;3^-\!\to\! 1^-) $
of \nuclide[20]{Ne}, \nuclide[24]{Mg}, and \nuclide[28]{Si} obtained with the MR-CDFT and MR-SCDFT calculations in comparison with experiment data~\cite{NNDC}. The unit of the transition strengths is $e^2\mathrm{fm}^4$.}
\begin{tabular}{l
    >{\centering\arraybackslash}p{1.25cm}
    >{\centering\arraybackslash}p{1.25cm}
    >{\centering\arraybackslash}p{1.25cm}
    >{\centering\arraybackslash}p{1.25cm}}
\hline
\hline
& \BEhead{2^+}{0^+}
& \BEhead{4^+}{2^+}
& \BEhead{3^-}{1^-}
& \BEhead{5^-}{3^-} \\
\hline
$^{20}$Ne \\ 
Exp. &65(3) &71(6)   & 161(26) &-\\
MR-CDFT &67  & 86 & 132 & 152\\
MR-SCDFT  &61 & 88  & 154 & 178 \\ 
\hline
$^{24}$Mg \\
Exp. &87(2) &147(14)  &- &132(58) \\   
MR-CDFT &86 &118  & 136 &157\\
MR-SCDFT  &88 &128  &167 &192\\ 
\hline
$^{28}$Si \\
Exp. &67(3) &83(9) &- &- \\
MR-CDFT &70 &105 &- &-\\
MR-SCDFT  &68 &111 &- &- \\ 
\hline\hline
\end{tabular}
\label{table2}
\end{table}

\begin{figure*}[t]
    \centering 
    \includegraphics[width=2\columnwidth]{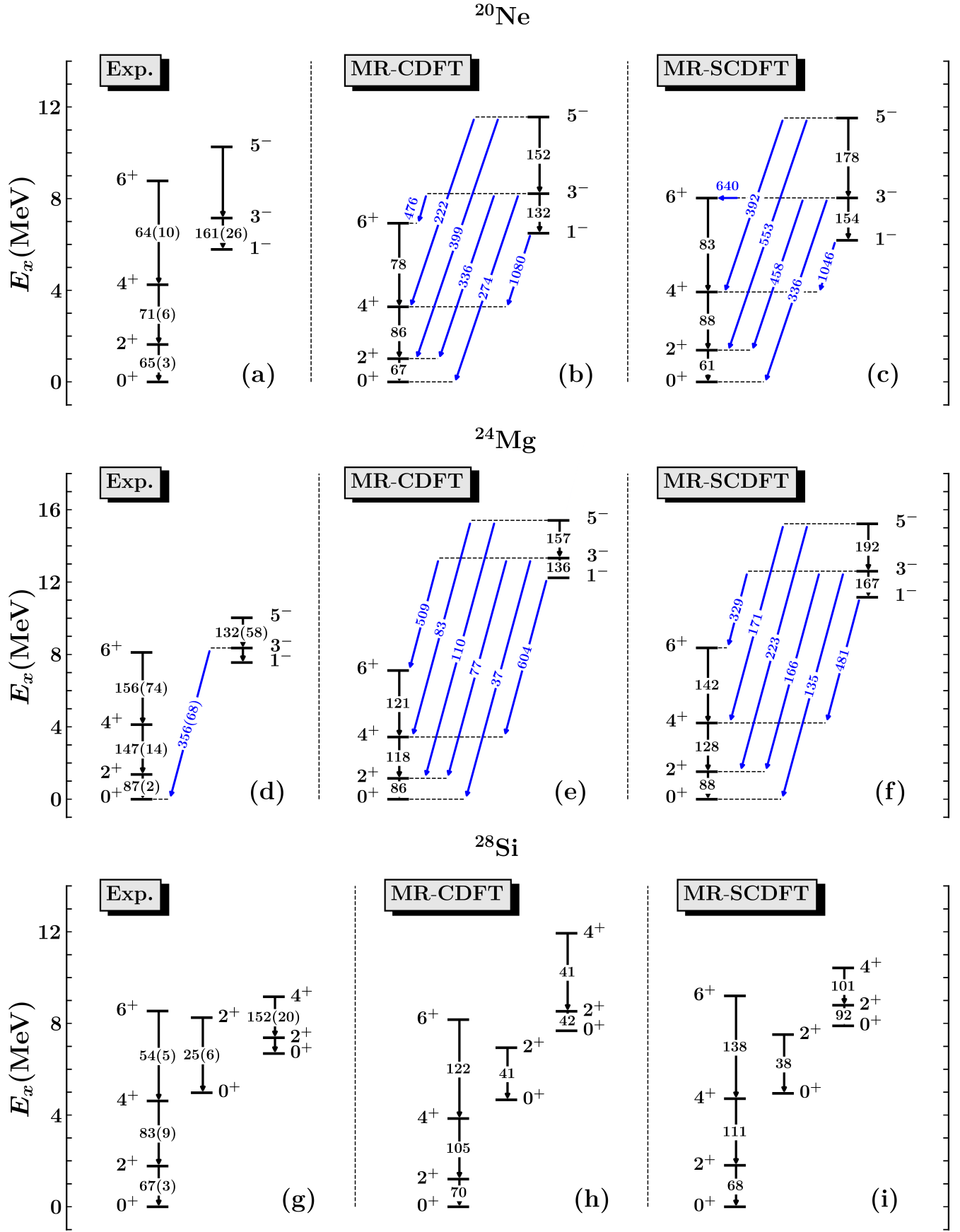}
    \caption{Same as Fig. 3 in the main text, but for \nuclide[20]{Ne}, \nuclide[24]{Mg}, and \nuclide[28]{Si}.}
    \label{fig4}
\end{figure*}

The results for the other $sd$-shell nuclei, $^{24}$Mg and $^{28}$Si, are 
summarized in Tables \ref{table1} and \ref{table2}. 
The low-lying spectra are also shown in Fig.~\ref{fig4}, together with the spectra for $^{20}$Ne. 

In Figs.~\ref{fig4} (e) and (f), one can notice that the calculated excitation energies of the $K^\pi=0^-$ band of \nuclide[24]{Mg} are overestimated. In Ref. \cite{Kanada2021PRC}, 
it was argued that 
there is 
pronounced $K$-mixing between the $K^\pi=0^-$ and $K^\pi=1^- $ configurations as evidenced 
by strong inter-band $E2$ strengths. This indicates a demand of the inclusion of triaxial 
degree of freedom, which is missing in the present calculation. 
Nevertheless, we notice that the energy of the $1^-$ band-head from MR-SCDFT is 
lower than the MR-CDFT value by about 1.1 MeV, 
in line with the importance of higher-order effects discussed in Ref.~\cite{Kanada2021PRC}.

The low-lying spectrum of $^{28}$Si is shown in Figs. ~\ref{fig4} (g), (h) and (i). 
The spectrum 
features three $  K^\pi=0^+  $ bands, that are commonly interpreted as the ground-state oblate rotational band, the second oblate $  \beta  $-vibrational band, and the third prolate shape-coexistence band~\cite{Frycz2024PRC}. Both MR-CDFT and MR-SCDFT overestimate the excitation energy of the band-head for the third $0^+$ band by roughly 1 MeV, that is consistent with its dominant 4-particle-4-hole (4p-4h) 
character revealed in shell-model studies~\cite{Frycz2024PRC}. 
Nevertheless, MR-SCDFT reproduces well the experimental band spacing of the third prolate band, 
whereas MR-CDFT fails to capture its rotational-like structure. 
This improvement indicates that the stochastic sampling 
in MR-SCDFT partially recover the essential features of the 4p-4h excitations. 
In contrast, the $  \beta  $-vibrational $2^+$ state is underestimated in both approaches, 
probably due to the lack the $  \gamma  $ degree of freedom as suggested in Ref. ~\cite{Frycz2024PRC}.

\section{Projection-Selection (PS) method}

In this work, we employ the projection-selection (PS) method to 
select important configurations for calculations based on a mixing of non-orthogonal configurations. 
The procedure of this method is as follows: 
\begin{enumerate}
\item[(\textrm{i})] All the configurations $  \ket{\Phi(\phi)}  $ are first sorted in ascending order of their projected energies, defined by the ratio of the diagonal components of the Hamiltonian and the norm kernels, $  {\cal H}^{J^\pi}(\phi,\phi)/{\cal N}^{J^\pi}(\phi,\phi)  $. This step requires $  \mathcal{O}(N_\phi)  $ operations, where $N_\phi$ is the number of configurations. Note that the energy ordering generally depends on the target angular momentum $  J  $.
\item[(\textrm{ii})] Starting from the lowest-energy configuration, a compact subspace is built iteratively. For each candidate configuration $  \ket{n+1}  $, one computes its overlap with the current subspace spanned by the already selected $n$ configurations:
\begin{align}
L(n,n+1) &= \frac{\langle n+1 | P^{(n)} | n+1 \rangle}{\langle n+1 | n+1 \rangle}
= \frac{\boldsymbol{\gamma}^{(n)\dagger} \bigl(\boldsymbol{S}^{(n)}\bigr)^{-1} \boldsymbol{\gamma}^{(n)}}{\langle n+1 | n+1 \rangle},
\label{eq:cutoff}
\end{align}
where $P^{(n)}$ is the projection operator onto the space $A_n$ 
spanned by the current selected $n$ configurations.  $S_{ij}^{(n)} = \langle i | j \rangle  $ and $  \gamma_i^{(n)} = \langle i | n+1 \rangle  $ with $i,j\in A_n$ are the matrix elements of the norm kernel $  {\cal N}^{J^\pi}(\phi,\phi')  $. If this overlap $  L(n,n+1)  $ is smaller than a pre-selected cutoff parameter $  L_c  $~\cite{Romero:2021PRC}, the new configuration is regarded as approximately orthogonal to the existing subspace and is added to it; otherwise, it is discarded. This screening process is repeated until all configurations have been examined, resulting in the final optimal subspace $  {\cal S}^{L_c}_N = \{\ket{1},\ket{2},\dots,\ket{N}\}_{L_c}  $. In the actual calculations shown in the main text, we take 
$L_c=0.83$. 
    \item [($\textrm{iii}$)]Because the off-diagonal kernels are important in the subsequent GCM calculation, in this step one adopts a more stringent screening procedure to further reduce the number of configurations. First, one iteratively eliminates configurations one by one: for each candidate in the current subspace $  {\cal S}^{L_c}_N $, one examines 
    its overlap $  L  $ with the remaining subspace $ {\cal S}^{L_c}_{N-1}  $. Any configuration satisfying $  L < L_{c_{\rm min}}  $ is discarded and the subspace is updated. This process is repeated until no further single configuration can be removed.
  Next, one considers pairs of configurations. For a candidate pair $  \{\ket{k}, \ket{l}\}  $ in the current subspace, one computes the $  2 \times 2  $ overlap matrix with respect to the remaining subspace $ {\cal S}^{L_c}_{N-2}  $:
\begin{align}
M_{kl} &= \frac{\langle k | P^{(n)} | l \rangle}{\sqrt{\langle k | k \rangle \langle l | l \rangle}}, 
\end{align}
The maximum $  L  $ value of this pair is given by the largest eigenvalue of the matrix $M$. If this eigenvalue is smaller than $  L_{c_{\rm min}}  $, the pair is discarded. After each removal the subspace is updated, and the procedure is repeated until no further pairs satisfy the removal criterion. This leads to the final subspace $ {\cal S}^{L_c}_{M}  $. Here, we stop at triplets of configurations, since eliminating larger clusters may result in a worse result of GCM calculations. In the actual calculations shown in the main text, we take 
$L_{c_{\rm min}}=0.01$. 
    \item [($\textrm{iv}$)] The norm and Hamiltonian kernels among the configurations retained in $ {\cal S}^{L_c}_{M}  $ are computed using the quantum-number projection (QNP) method, and the Hill-Wheeler-Griffin (HWG) equation is solved to obtain the final many-body states.
\end{enumerate}

\clearpage
\onecolumn
\bibliographystyle{elsarticle-num}
\bibliography{ref}